# Model Channel Ion Currents in NaCl – SPC/E Solution with Applied–Field Molecular Dynamics

Paul S. Crozier*, Douglas Henderson[†], Richard L. Rowley*, and David D. Busath[‡]

*Department of Chemical Engineering [†]Department of Chemistry and Biochemistry [‡]Department of Zoology and Center for Neuroscience Brigham Young University Provo, Utah 84602

**ABSTRACT**

Using periodic boundary conditions and a constant applied field, we have simulated current flow through an 8.125 Å internal diameter, rigid, atomistic channel with polar walls in a rigid membrane using explicit ions and SPC/E water. Channel and bath currents were computed from ten 10-ns trajectories for each of 10 different conditions of concentration and applied voltage. An electric field was applied uniformly throughout the system to all mobile atoms. On average, the resultant net electric field falls primarily across the membrane channel, as expected for two conductive baths separated by a membrane capacitance. The channel is rarely occupied by more than one ion. Current-voltage relations are concentration-dependent and superlinear at high concentrations.

## INTRODUCTION

Many molecular dynamics (MD) studies have been performed with atomistic models of ion channels during the past two decades (for exemplary reviews, see Roux and Karplus, 1994; Forrest and Sansom, 2000). These have focused primarily on the structures of the channels and the energetics and dynamics of their contents, which consist of explicit water molecules and one or a few ions in the channel with no applied potential. In the past few years, this type of study has dramatically intensified (Allen et al., 1999; Guidoni et al., 1999; Roux and MacKinnon, 1999; Allen et al., 2000; Aqvist and Luzhkov, 2000; Berneche and Roux, 2000; Capener et al., 2000; Guidoni et al., 2000; Hu et al., 2000; Roux et al., 2000; Shrivastava and Sansom, 2000) with the discovery of the crystal structure of a bacterial potassium channel (Doyle et al., 1998), which is expected to serve as a prototype for the structures of voltage-gated channels.

These computations have drawn attention to the structure and reduced diffusion coefficients of water molecules in the confined space and have demonstrated some of the energetic components responsible for ion selectivity in biological channels, but generally do not attempt to simulate ion flow in the aqueous baths outside the channel or the process of channel entry and crossing.

The equilibrium properties of ions in channels have also been studied with statistical mechanics and simulation approaches using simplified channel models (Vlachy and McQuarrie, 1986; Sorensen and Sloth, 1992; Lozada-Cassou et al., 1996; Lynden-Bell and Rasaiah, 1996; Hartnig et al., 1998; Lo et al., 1998; Roux, 1999; Tang et al., 2001; Tang et al., In press). For instance, based on the mean spherical approximation, the excess chemical potential for binding divalent ions relative to monovalent ions in a confined space with constrained charges helps explain the selectivity of a voltage-gated calcium channel for $Ca^{++}$ over $Na^+$ (Nonner et al., 2000). The result, which identifies space/charge competition as the mechanism of binding selectivity, has been confirmed by canonical ensemble Monte Carlo simulations used to determine the distribution of ions between a cylindrical bath surrounding a periodically infinite "channel" containing confined negative charges

(Boda et al., 2000; Boda et al., in press; see also Golding et al., 2000).

Efforts to examine the flow of ions into and through channels have been carried out using Nernst-Planck (NP) (Levitt, 1978; Levitt, 1982; Levitt, 1987; Sancho and Martínez, 1991), Poisson-Nernst-Planck (PNP) simulations (Peskoff and Bers, 1988; Chen et al., 1997a,b; Chen et al., 1999; Kurnikova et al., 1999; Hollerbach et al., 2000, Cardenas et al., 2000), and Brownian dynamics (BD) (Jakobsson and Chiu, 1987; Chiu and Jakobsson, 1989; Bek and Jakobsson, 1994; Chung et al., 1998; Chung et al., 1999; Im et al., 2000; Corry et al., 2000a; Corry et al., 2001). These simulations treat ions as point charges (NP, PNP) or spheres (BD), and the water as a viscous continuum dielectric in order to speed simulation processing. The BD simulations are superior to the NP and PNP simulations for narrow pores because the volumes of the ions are considered explicitly (Moy et al., 2000; Corry et al., 2000b). From studies of current-voltage-concentration relationships with BD, it is clear that substantial radial dipole potentials are required to offset the dielectric boundary effects in order for ions to enter channels like the nicotinic acetylcholine receptor (NAChR) (Chung et al., 1998), that the dipoles in pore lining of the KcsA potassium channel allow the observed multiple occupancy and permeability of the channel (Chung et al., 1999), and that the constriction zone in OpmF porin channels presents an energy barrier that is responsible, rather than selective vestibule occupancy, for the observed channel selectivity (Im et al., 2000).

Perhaps more importantly from a methodological point of view, these BD studies have begun to address a central issue about boundary conditions: how to treat the connection to the essentially infinite bath and membrane found in experimental conditions. Im et al. (2000) use grand canonical Monte Carlo (GCMC) steps in two thin slabs of solution 15 Å from the membrane surfaces to maintain constant chemical potential in the baths and BD steps to simulate ion flow through the baths and channel. This approach demonstrates a key issue: the ion occupancy of the volumette near the entry and exit of the channel fluctuates considerably and is a Poisson distributed random variable if interactions between particles are neglected (Roux, 1999; Im et al., 2000).

Furthermore, the dynamics of ion permeation have been illuminated and shown in some cases to differ from expectations based on preconceptions about free energy profiles and transport over energy barriers. For instance, potassium entry into the cytoplasmic end of a smooth-walled model of the KcsA potassium channel is only weakly (rather than linearly) dependent on cytoplasmic $[K^+]$ and is strongly dependent on membrane potential (Chung et al., 1999). Both of these results are counterintuitive from the point of view of near-equilibrium permeation theory (i.e. rate theory; Hille, 1992) because collisions of cytoplasmic ions with the channel are expected to rise linearly with concentration (Läuger, 1976) and to be relatively independent of applied potential due to the conductive nature of the bath (Andersen, 1983a).

BD simulations may suffer from neglect of the volume and molecular polarization of water molecules, especially as they mediate interactions between ions within the channel, between ions and the channel walls, and between ions in the channel and ions near the channel entry or exit. It is therefore desirable to consider the effects of solvent and ion momentum explicitly using classical MD. This requires a small system size, but is shown here to be feasible with periodic boundary methods that allow continuous flow without ion repositioning. We refer here to this particular form of non-equilibrium MD (NEMD) as applied-field MD.

The model channel and membrane system that we use is simple, consisting of a rigid atomic pore with polar walls (i.e. partial charges on the pore atoms) and internal diameter similar to that of NAChR (Hille, 1992) embedded in a rigid, uncharged membrane. The rigid membrane helps prevent accumulation of momentum along the axis of channel flow and enhances computational efficiency, which, together with the small size of the system and the $P^3M$ Ewald sum electrostatics, made it feasible to simulate a period sufficient to measure current flow.

## COMPUTATIONAL METHODS

### Model System

Applied-field NVT NEMD simulations were performed using a $25 \times 25 \times 55$ Å (in the *x*-, *y*-, and *z*-directions respectively) simulation box with periodic boundary conditions in all three directions. Rigid, fixed-in-space, model membrane walls consisted of neutral Lennard-Jones (LJ) spheres placed on square lattices at $z = 15$ Å and $z = 40$ Å. Center-to-center spacing of the LJ spheres in the $x$- and $y$-directions was set at 2.5 Å, and the LJ parameters for each were set at $\sigma = 2.5$ Å and $\varepsilon/k = 60$ K, with cross interactions between the mobile particles calculated using standard Lorentz-Berthelot (LB) rules. A 4-sphere × 4-sphere section centered at $x = y = 12.5$ Å in each 10-member × 10-member wall was removed to form the entrance to the channel structure.

The model channel structure was formed using eleven twenty-member rings of LJ spheres with the same parameters as those assigned to the membrane spheres. Rings were given a center-to-center diameter of 10.625 Å, which yields an internal diameter for the channel of 8.125 Å (after subtracting two atomic radii, or $\sigma$). In addition to the LJ parameters, each channel sphere was assigned a partial charge of $-0.5\ e$, $+0.5\ e$, $-0.35\ e$, or $+0.35\ e$ in a repeating pattern around each identical twenty-member ring ($e$ being the elementary charge). The partial charges were designed to simulate those commonly used for the peptide units in proteins (Brooks et al., 1983) and approximately simulate the polarity of a backbone-lined channel such as the gramicidin channel or the P region of the potassium channel. These spheres were also held rigid at even spacing around the perimeter of each ring, with each ring centered at $x = y = 12.5$ Å and placed at 2.5 Å intervals along the $z$-axis from $z = 15$ Å to $z = 40$ Å, forming a tube connecting the two membrane walls. Each successive ring was rotated 9º about the $z$-axis relative to the previous ring in order to produce a helical pattern of charge distribution along the tube as shown in Figure 1. The positioning of the membrane and channel spheres rendered the walls impermeable to the mobile atoms. The remainder of the volume of the simulation cell was accessible to the mobile particles making up the aqueous electrolyte solution.

The electrolyte solution consisted of a combination of SPC/E water molecules (Berendsen et al., 1987), $Na^+$ ions, and $Cl^-$ ions. SPC/E water was used rather than TIP3P water because in preliminary simulations with 1.0 M NaCl in TIP3P water we found that the ions had an anomously strong tendency to form ion pairs and clusters. The LJ and coulombic interaction parameters for the model water molecules and model ions were those used by Spohr (1999), and are repeated in Table 1 for convenience. Note that the cross parameters differ from those that would be obtained from the LB rules.

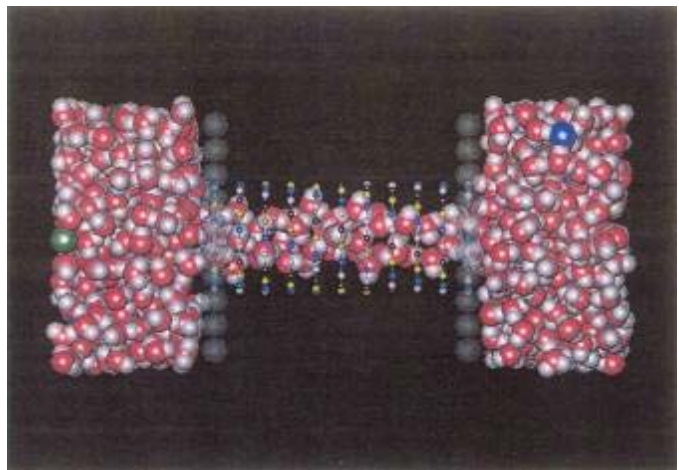

**Figure 1.** Snapshot of the 1 M NaCl in SPC/E water simulated system showing the channel and membrane structure. Sodium and chloride ions are the large green and blue spheres, respectively. Neutral membrane atoms are drawn as transparent light blue spheres, while the charged atoms comprising the channel walls are depicted as small spheres for ease in viewing channel contents. The small black, blue, yellow, and white channel wall spheres carry charges of –0.5, -0.35, 0.35, and 0.5 $e$ respectively. True system and species dimensions are given in the text. This image and those in figures 12 and 13 were rendered using VMD (developed by the Theoretical Biophysics Group in the Beckman Institute for Advanced Science and Technology at the University of Illinois at Urbana-Champaign), (Humphrey et al., 1996).

**Simulation Procedure and Details**

Ten sets of ten simulations were performed using the model system described above.

|  | $\sigma$ (Å) | $\varepsilon/k$ (K) |
|---|---|---|
| **O-O** | 3.169 | 78.187 |
| **O-$Na^+$** | 2.876 | 62.724 |
| **O-$Cl^-$** | 3.250 | 62.724 |
| **$Na^+$-$Na^+$** | 2.730 | 43.001 |
| **$Na^+$-$Cl^-$** | 3.870 | 20.512 |
| **$Cl^-$-$Cl^-$** | 4.860 | 20.186 |

**Table 1.** Electrolyte solution Lennard-Jones parameters.

Systems containing a nominal ion concentration of 0.5, 1 or 2 M were each tested at 0.55, 1.1 or 2.2 V externally applied potential, as well as a tenth set at 0.5 M, and 0.0 V. The exact numbers for the nominal concentrations were 608 water molecules, 4 $Na^+$ ions, and 4 $Cl^-$ ions for the 0.5 M case; 600 water molecules, 8 $Na^+$ ions, and 8 $Cl^-$ ions for the 1 M case; and 584 water molecules, 16 $Na^+$ ions, and 16 $Cl^-$ ions for the 2 M case. These systems were close to experimental liquid densities and, as shown below, had reasonable ion and water mobilities.

Each of the 100 runs consisted of 1 ns of equilibration time followed by 10 ns of data collection using a time step of 2.5 fs. Each run was performed on a single CPU of a 64-node SGI Origin 2000 supercomputer and required approximately two weeks of CPU time. Mobile particle positions were stored at 2.5 ps intervals (every 1000 time steps) for analysis by a post-processor program.

An electric field producing the membrane potential was uniformly applied in the *z*-direction to all mobile particles in the simulation cell, whether in the bath or in the channel, producing the specified potential drops across the 55 Å simulation cell. Because of the external electric field we expect to see some charge build-up in the form of an electrochemical double layer at both membrane walls. It will be shown later that as an ensemble average, this double layer does indeed form in accordance with expectations from the membrane capacitance, neutralizes the electric field in the reservoir region, and, under opposition from channel water, magnifies the field across the membrane. The combination of the applied field and the response of the mobile particles yields, approximately and on average, the expected constant electric field across the membrane and zero field in the conductive baths, resulting in a membrane potential equal to the drop in the applied potential across the unit cell. The average state generally consists of an ion-free channel; it will also be shown below that during current flow the membrane potential changes.

Previous work (Torrie and Valleau, 1980; Eck and Spohr, 1996; Crozier et al., 2000b) shows that proper simulation of the electrochemical double layer requires adequate representation of long-range electrostatic interactions, including those acting beyond the dimensions of the primary simulation cell. Several methods have been developed for the estimation of these long-range forces, including the charged sheets method (Torrie and Valleau, 1980; Boda et al., 1998), the Ewald sum method (Parry, 1975), and mesh Ewald methods (Hockney and Eastwood, 1988; Darden et al., 1993; Essmann et al., 1995). The charged sheets method has been shown to be inadequate for our purposes (Crozier et al., 2000b), and the standard Ewald sum method is far too computationally demanding (Crozier et al., 2000a). We use the particle-particle/particle-mesh ($P^3M$) method because of its demonstrated flexibility and superiority to other mesh Ewald-methods (Deserno and Holm, 1998a). We follow the implementation recommendations of Deserno and Holm (1998a, 1998b), and refer to their excellent discussion of optimization and error minimization for mesh-Ewald calculations (1998b).

$P^3M$ implementation details are as follows. A seventh-order charge assignment function was used along with a 16-point × 16-point × 64-point grid in the *x*-, *y*-, and *z*-directions, respectively. All LJ interactions and real-space coulombic interactions were truncated at $r_{cut}$ = 10 Å. The reciprocal-space portion of the Coulombic interactions was determined by 1) assignment of the charges to the mesh according to the seventh-order charge assignment function, 2) transformation of the charged mesh to Fourier space using a fast Fourier transform, 3) multiplication by the optimal influence function to determine the potential at each mesh point, 4) *i***k** differentiation in each direction to find the respective electric fields, 5) inverse fast Fourier transform back to real space for fields in all three directions, 6) assignment of the mesh-based electric fields back to the particles according to the same seventh-order charge assignment function, and 7) computation of the reciprocal-space force contribution on each particle given the electric field and the charge on each particle. For our model system, $\alpha$, the parameter that divides the $P^3M$ calculation into real- and reciprocal-space contributions, was set at a constant value of 0.3028 $Å^{-1}$ according to the optimization scheme of Deserno and Holm (1998b). The optimal influence function was computed only once for each run (at the beginning).

Gaussian bond and temperature constraints were used (Edberg, et al. 1986; Rowley and Ely, 1981), with feedback correction to remove numerical drift error. In all cases, the system temperature was maintained at 25º C. A fourth-order Gear predictor-corrector integration scheme was used to integrate the equations of motion.

Channel (and bath) currents were computed from net charge displacements during intervals of $\Delta t=2.5$ ps as:

$$i = \frac{\sum_j q_j \Delta z_j}{\Delta t L} \quad [1]$$

where $i$ is the calculated current, $q_j$ is the charge on ion $j$, $\Delta z_j$ is the net –displacement of ion $j$ within the channel (or bath) during the interval $\Delta t$, and $L$ is the channel (or bath) length. The sum is over all ions appearing in the channel (or bath) during the interval, including interpolations for charges moving from one region to the other during $\Delta t$. Small charge movements due to rotations of water molecules were neglected.

**Results**

As implied above, the applied field is oriented such that it drives the mobile positively charged atoms towards higher values of z, and negatively charged atoms towards lower values of z, and rotates water molecules to orient their dipole vectors parallel to the applied field. In the plots used here, the z-position is measured from the left-hand periodic boundary. The membrane is centered in the simulation cell between z values of 15 and 40 Å.

- *Ion Trajectories*

The z-coordinates of all the ions in the system are shown as a function of time during the simulation period for one of the conditions tested, (2 M NaCl, 0.55 V total applied potential) in Figure 2. The ten 10-ns runs have been concatenated into one trace. Thus some ion passage trajectories that appear to terminate or initiate abruptly in mid channel really represent events occurring at the concatenation boundary. Because of the large applied potential, the ion motions along the z direction within the channel are quite uniform with minor fluctuations. Nine complete $Na^+$ passages and one $Cl^-$ passage (starting at 50 ns) can be observed, with the $Cl^-$ passing in the opposite direction (from high to low z), as expected. The passages appear to occur randomly in time, as expected for a stochastic process. Occasional visits of $Cl^-$ ions at the negatively polarized interface (40-43 Å) and of $Na^+$ ions at the positively polarized interface (12-15 Å) can be identified in Figure 2. These partly represent capacitive charge, but in many cases they are due to partial channel entries, especially where the ions get closer than $\sigma$ (~2.5 Å) to the membrane sphere centers at z=15 Å and z=40 Å. In the baths (z=0-10 Å and z=45-55 Å), the ions are uniformly distributed. For our model channel, the time-course of the channel current during a single ion passage is not rectangular, both because of random fluctuations and because the velocity tends to increase as the ion approaches the exit. This is seen in Figure 3, which shows the average velocity of Na+ as a function of z-position at each of the three applied potentials utilized for 2 M NaCl. The velocity starts at 2-3 m/s for 0.55 and 1.1 V, 4-5 m/s for 2.2 V. For the lowest voltage, it remains relatively constant throughout the passage, until it nears the exit (38 Å), at which point it undergoes an abrupt increase up to ~16 m/s (39 Å) and then falls back to <5 m/s as the ion reaches the outer ring of channel atoms and membrane atoms (40 Å), and to near 0 m/s as the ion enters the bath. At the higher applied fields, the ion gains momentum as it passes through the channel.

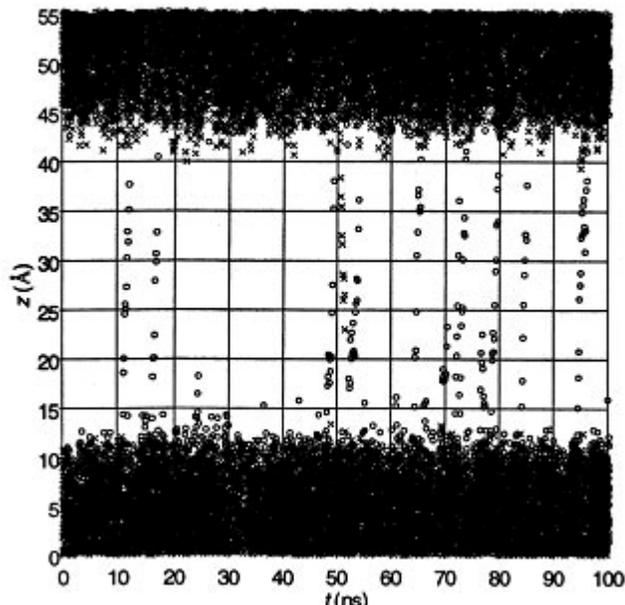

**Figure 2.** $Na^+$ (O) and $Cl^-$ (×) ion z-positions as a function of time for the case of 2 M NaCl with 0.55 V applied potential. Symbols appear at 100 ps intervals. The ten runs of 10 ns each are plotted consecutively for a total of 100 ns of simulation time.

Statistics taken from the trajectories for all of the 10 cases are given in Table II. The first two columns contain the independent variables for the 10 cases. The average time for passage through an entry region (including failed entries) extending from 12.5-17.5 Å, through the interior from 17.5-37.5 Å, and through an exit region of the channel from 37.5-42.5 Å are given in three rows per cell in the third column. These are calculated from the inverse of the mean ion velocities in these regions. The passage time for the four-fold longer interior step is 0.32 – 2.24 ns, generally less than four times the entry step passage time and much more than four times the exit step passage time, contrary to what one would expect if ion velocity were constant throughout the passage. The long average entry passage time reflects the fact that many ions penetrate only a short distance into the channel and then return out of the channel, reducing the population average forward velocity. The short average exit passage time is due to rapid acceleration of the ion as it approaches the exit. Passage times through all three regions decrease with increasing applied potential but are independent (to within statistical errors) of bath concentration.

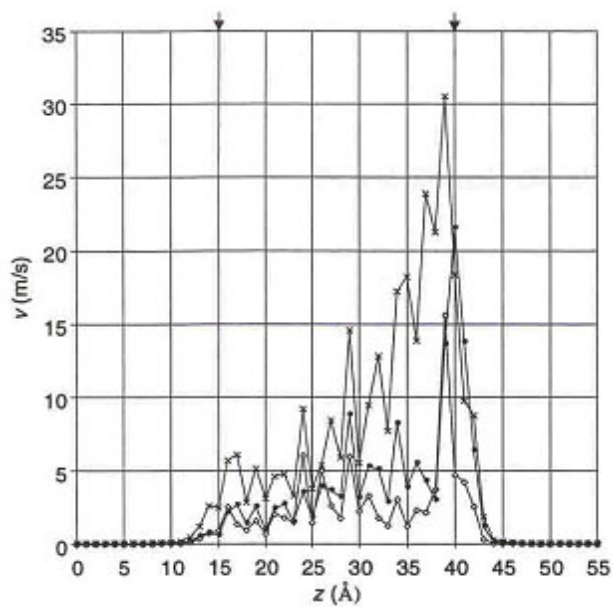

**Figure 3.** Average net velocity of sodium ions as a function of $z$-coordinate. Data is shown for the 2 M NaCl cases at 0.55 V (◊), 1.1 V (●), and 2.2 V (×). The arrows in this and subsequent Figures (5-8) mark the entry and exit of the channel.

The channel and bulk currents calculated from equation 1 are given in the fourth column of Table 2. For reference, 1 pA represents 0.625 complete channel crossings per 100 ns simulation time. Although the standard deviations for the currents are not presented, we note that they were much smaller for the channel currents than for the bulk currents. However, both have an uncertainty due to counting errors associated with the Poisson process of channel crossings, which can be deduced from the square root of the number of observed crossings. For the lower concentration, the counts limit our certainty of passage frequency to an order of magnitude, so we do not focus on detailed patterns in the currents. Rather, we simply note that the channel and bulk currents generally agree to within a few pA of statistical fluctuation, as expected from Kirchhoff's current law (Horowitz and Hill, 1982), and they generally increase with applied potential and concentration.

The next 5 columns of Table 2 present the average time fractions for the occupancy states for the main channel region (15-40 Å). Although the accuracies of these numbers are limited only by the frequency with which we saved trajectory frames, their usefulness for describing the steady state probabilities of occupancy depends on the underlying statistics of the random processes of ion entry and exit. Because in the best case (2 M, 2.2 V) we only observed ~62 crossings, the accuracy of such Poisson statistics is only of the order of 10%. We therefore show them only to suggest trends, but we feel that detailed kinetic analysis of the results (in terms of discrete step models, for instance) is unwarranted with this data set. The channel is primarily occupied by water only (no ions) or by one $Na^+$. Occasional occupancy by $Cl^-$, $Na^+$ and $Cl^-$, or two $Na^+$ ions are noted, especially at 2 M and 2.2 V, but it is obvious that the model channel is primarily selective for $Na^+$ and is a single- rather than multiple-occupancy channel. Double $Cl^-$ and triple occupancy (by any ions) were never observed.

For narrow cylindrical channels, for example gramicidin, saturation in the current-voltage relation at high voltages (Andersen, 1983a) suggests that the rate of ion entry is essentially voltage-independent, presumably because the electric field in the conductive bath is negligible; but translocation (and, to a lesser extent, exit from the channel) are expected to increase substantially with voltage. This combination leads to the prediction that channel occupancy should decrease as applied potential is increased. We note an opposite trend here, at least for differences between 0.55 and 1.1 V. $Na^+$ occupancy ranges from .021 to .302, increasing with concentration as expected (because the entry rate is expected to be proportional to the bath concentration), but also increasing with voltage.

This implies that the effective entry rate is voltage dependent, more so than translocation and exit from the channel at these voltages.

The average ionic current through the channel is shown as a function of total applied potential in

Figure 4. As expected, the current increases with applied potential and with concentration of the bathing ion. Although the current-voltage (I-V) relationship appears to be superlinear in 1 M and 2 M NaCl and sigmoidal in 0.5 M NaCl, the large size of the error bars prevent detailed assertions about the I-V shapes. Therefore, we have chosen just to draw a spline curve through the points as an eye guide, rather than to fit a simplified model function. Nevertheless, there appears to be a progression in shape from sub- to superlinear with increasing concentration.

| Conc. (mol/L) | V (volts) | t (ns) entry pass. exit | i (pA) chan bulk | P(0) | P(Na$^+$) | P(Cl$^-$) | P(Na$^+$+Cl$^-$) | P(2 Na$^+$) |
|---|---|---|---|---|---|---|---|---|
| | 0.0 | N/A | 0.00 -0.49 | 0.996 | 0.004 | 0 | 0 | 0 |
| | 0.55 | 2.15 1.09 0.18 | 2.60 1.74 | 0.979 | 0.021 | 0 | 0 | 0 |
| | 1.1 | 0.24 0.65 0.05 | 19.7 21.1 | 0.908 | 0.091 | 0.001 | 2E-5 | 0 |
| | 2.2 | 0.20 0.33 0.02 | 21.6 20.8 | 0.931 | 0.046 | 0.023 | 0 | 0 |
| | 0.55 | 0.76 2.24 0.12 | 8.52 7.90 | 0.872 | 0.126 | 0.002 | 0 | 0 |
| | 1.1 | 0.30 0.80 0.05 | 24.7 19.0 | 0.859 | 0.139 | 0.002 | 0 | 0 |
| | 2.2 | 0.18 0.37 0.03 | 68.1 69.1 | 0.800 | 0.174 | 0.019 | 0.007 | 0 |
| | 0.55 | 0.63 1.12 0.12 | 18.1 20.3 | 0.852 | 0.137 | 0.010 | 0.001 | 0 |
| 2 | 1.1 | 0.57 0.73 0.07 | 35.4 38.1 | 0.792 | 0.206 | 0.001 | 0.001 | 7E-5 |
| | 2.2 | 0.20 0.32 0.04 | 99.5 98.8 | 0.740 | 0.203 | 0.047 | 0.007 | 0.003 |

**Table 2.** Mean entry (12.5 < $z$ < 17.5 Å), passage (17.5 < $z$ < 37.5 Å), and exit (37.5 < $z$ < 42.5 Å) times for Na+ ions, average current as measured in the channel region (15 < $z$ < 40 Å) and bulk reservoir region, and occupancy probabilities for the ten cases tested. Probabilities are defined as the total time that the system was found in a given state divided by the total simulation time, where the occupancy states were: no ions, one Na$^+$ ion, one Cl$^-$ ion, one Na$^+$ and one Cl$^-$, or two Na$^+$ ions in the channel region. On no occasion did more than one Cl$^-$ or more than two ions occupy the channel.

- *Ion/water structure*

The current flow through the channel is related to the charge structure throughout the system, i.e. the positions of ions and water dipoles. As a description of this structure, Figure 5 gives the average localized density of each of the four charged species as a function of z for the 100 ns simulation at 0.5 M and 1.1 V. For each point, the local density (in units of moles/liter) is calculated from the average occupancy of a narrow slab centered at that point. In the region of the bulk and the interface, the volume accessible to the mobile atom centers is taken as that of a square slab 25 Å on a side (parallel to the x-y plane) and of thickness (in the z dimension) of 0.025 Å. In the region of the channel (15 < z < 40 Å), the cross sectional area of a right cylinder approximately in contact with the van der Waals surface of the channel atoms, 51.85 Å$^2$, is used to define the volume accessible to the mobile atoms in the channel. This is an arbitrary volume, because the different atom species differ in their van der Waals parameters, but serves to normalize the density calculation for the sake of comparison between channel and bulk.

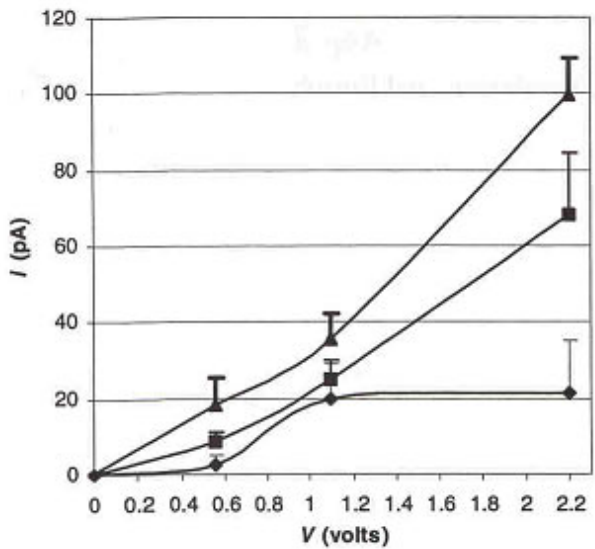

**Figure 4.** Current as a function of externally applied potential and ion concentration. The diamonds (♦), squares (■), and triangles (▲) represent the 0.5 M, 1 M and 2 M cases respectively. Error bars represent +1 standard deviation of the calculated currents. Solid curves are eyeguide splines drawn manually.

The ion and water atom densities approach a bulk-like uniform distribution by ~7 Å from the bilayer. This uniformity is one success of the applied field NEMD method (Crozier et al., 2001). In spite of the applied field in the bulk region, the densities are, on average, constant and the net electric field negligible (see below) as expected for a conducting medium.

During this simulation, the channel was occupied only by water molecules 90.8% of the time and by Na$^+$ (and water) 9.1% of the time (Table II), so the profile primarily reflects the state of the unoccupied channel. [Na$^+$] and [Cl$^-$], plotted at ten times their actual values, are 0.5 M in the bulk, but [Na$^+$] rises above 0.5 M near the membrane on the left (6 < z < 8 Å) whereas [Cl$^-$] falls below. In contrast, [Cl$^-$] rises slightly above and [Na$^+$] falls below on the right (45 < z < 47 Å). The asymmetry between cation and anion concentration continues up to the channel entrance, defined as the planes of the membrane atom centers at 15 and 40 Å. Of course, the excess average of cations on the left and anions on the right cannot be detected in any single snapshot, but represents a probability density constituting the capacitative charge on the membrane. In this case, the average net charge is +1.29×10$^{-6}$ coul/cm$^2$ (0.54 e) on the right and −1.62 ×10$^{-6}$ coul/cm$^2$ (-0.63 e) on the left, corresponding to a specific capacitance, Q/VA, of 2.74 µF/cm$^2$ for the membrane. This is nearly 10-fold higher than would be computed from $\varepsilon_0/d$ = 0.30 µF/cm$^2$, where $\varepsilon_0$ is the permitivity of free space and the membrane is assigned a thickness of d=30 Å (which approximately takes into account the membrane interface atom radii). This is also higher than is usually observed for lipid membranes

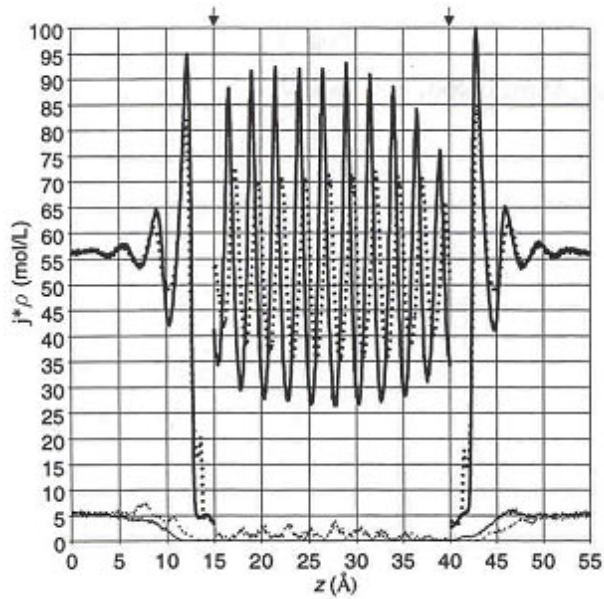

**Figure 5.** Average density of the various species as a function of z for the 0.5 M and 1.1 V applied potential case. The SPC/E water density is divided into O (solid bold line) and H (dotted bold line) species and is seen above the chloride ion (solid line) and sodium ion (dotted line) density distributions. For improved viewing, ion densities were multiplied by different factors, namely j=1.0 (O), 0.5 (H), 10.0 (Na$^+$), and 10.0 (Cl$^-$). Moreover for the channel region (15 Å < z < 40 Å), all densities were multiplied by an additional factor of twelve, because only roughly 8.3% of the volume in that region was accessible to the particles.

experimentally (~1 μF/cm$^2$). The discrepancy is probably due to the water in the channel.

In Figure 5, the density of the SPC/E water near the two boundaries of the system (z = 0 and 55 Å) is 56.4 M, approximately that of bulk waterAlthough there is no reason to expect the SPC/E model of water to give reasonable pressures in this NVT simulation, mobility of the ions and water were reasonable, with the average water diffusion coefficient in the bulk compartment (specifically 0 < z < 5 Å and 50 < z < 55 Å) being computed as 2.3×10$^{-5}$ cm$^2$/s, that of Na$^+$ as 1.3×10$^{-5}$ cm$^2$/s, and that of Cl$^-$ as 1.1×10$^{-5}$ cm$^2$/s. Three layers of water ordering can be observed with increasing structure
near the membrane surface, seen as peaks in oxygen density at 5, 9, and 12 Å on the left. Similar peaks occur on the right. The water structure may be important in the overall potential profile for this system, but may not be significant to biological systems where the lipid head group structure is much more complex and the head groups are mobile.

The water oxygen atoms in the channel have an average density near that in the bulk. They are clearly structured in 10 layers, corresponding to the eleven rings of atoms forming the channel wall, and oriented by the applied potential. The Na$^+$ density in the channel also shows peaks. These are located at the oxygen minima and suggest a steric preference for more vacant locations. As will be shown below, strong interactions between water and the channel walls are only modestly disrupted by Na$^+$ in the channel. Electrostatic potential

The contributions of the particles of the system and the applied potential to the net electrical potential along the axis of the channel was computed from the ensemble average charge distribution. In doing this, we took care to avoid delta function contributions to the potential due to close proximity of charged particles to the test point by performing the calculation strictly in reciprocal space, setting $r_{cut}$ to 2 Å, re-adjusting the tuning parameter, α, and excluding the real-space contributions. This method yielded the same result as was obtained by direct integration of charge along z (based on the Poisson equation) for a system with a bilayer but no channel (i.e. homogeneous in two of the three dimensions). The total potential drop for the four voltages studied is shown in Figure 6. The figure is dominated by the no-ion occupancy state, as the channel is mostly ion-free. As can be seen, there is very little potential drop in the bath for the ion-free channel state and the total voltage drop of 1.1 V occurs across the membrane.

To help understand the origin of this electrostatic potential, it is useful to examine how the applied potential orients the water molecules in the system. This is shown in Table 3, which gives the total ensemble average cosine of the angle between the water dipole and the z axis for each of four different regions. Column I is the central region of the bath near the periodic boundaries (z<5 or z>50 Å), column II is near the left wall (5<z<15 Å), column III is in the channel (15<z<40 Å), and column IV is near the right wall (40<z<50 Å). This statistic would be zero for fully randomized water orientations. It should be noted that the orientation increases with applied potential near the walls and in the channel, and is negligible everywhere when no potential is applied and at all voltages in the central region. The slight reductions in channel water orientation at higher bath ion concentrations may reflect increased disruption of the channel waters by passing ions.

The electrostatic potential can be decomposed into water and ion contributions that can be understood in terms of two major dipoles: the channel water dipole and the capacitative charge dipole. The water molecules in the channel are oriented by the net potential (Figure 5, Table 3) with the negative end of the dipole towards the channel entry at z=15 Å. The ion distribution at the membrane water interface forms the capacitative charge, which constitutes a macroscopic dipole oriented in the opposite direction. The channel water produces an increased local membrane capacitance and affects the ion density in the bath near the channel openings.

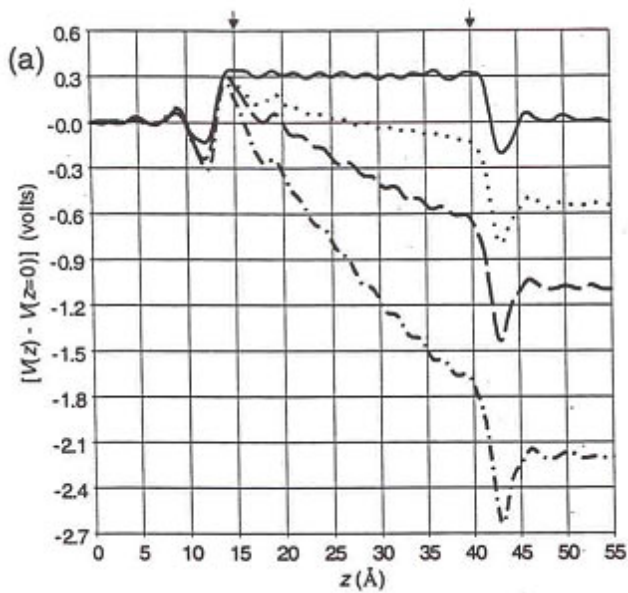

**Figure 6.** Comparison of the total electrostatic potential as a function of *z* for the 0.5 M systems with 0.0 V (solid line), 0.55 V (dotted line), 1.1 V (dashed line), and 2.2 V (dash-dot line) applied potential. In each case the potentials were calculated using a test charge positioned on the z axis for the average charge distribution from the entire simulation.

| Conc. (mol/L) | V (volts) | _cos(θ)_ | | | |
|---|---|---|---|---|---|
| | | I | II | III | IV |
| 0.5 | 0.0 | 0.00 | -0.02 | -0.08 | 0.00 |
| | 0.55 | 0.00 | 0.03 | 0.66 | 0.06 |
| | 1.1 | 0.02 | 0.04 | 0.76 | 0.07 |
| | 2.2 | 0.01 | 0.06 | 0.83 | 0.08 |
| 1 | 0.55 | 0.00 | 0.02 | 0.63 | 0.05 |
| | 1.1 | 0.01 | 0.04 | 0.75 | 0.06 |
| | 2.2 | 0.01 | 0.05 | 0.81 | 0.07 |
| 2 | 0.55 | 0.00 | 0.02 | 0.62 | 0.04 |
| | 1.1 | 0.00 | 0.02 | 0.74 | 0.05 |
| | 2.2 | 0.01 | 0.04 | 0.79 | 0.06 |

**Table 3.** Ensemble average water orientation for region I near the boundaries, II near the left wall, III in the channel, and IV near the right wall. See text for details.

The total potential drop across the membrane has an approximately linear profile except at the ends of the channel, where the contribution of the water molecules results from oriented water dipoles in the structured layers near the interface and contribute a significant biphasic potential. This has the shape of the potential expected from a combination an infinite sheet of surface charge and an infinite sheet of dipoles. In the present case, the potential is due strictly to the water molecules and ions because the membrane atoms are all neutral. The total potential drop occurs primarily in the channel region at the four applied potentials of 0.0, 0.55, 1.1, and 2.2 V as shown in Figure 6.

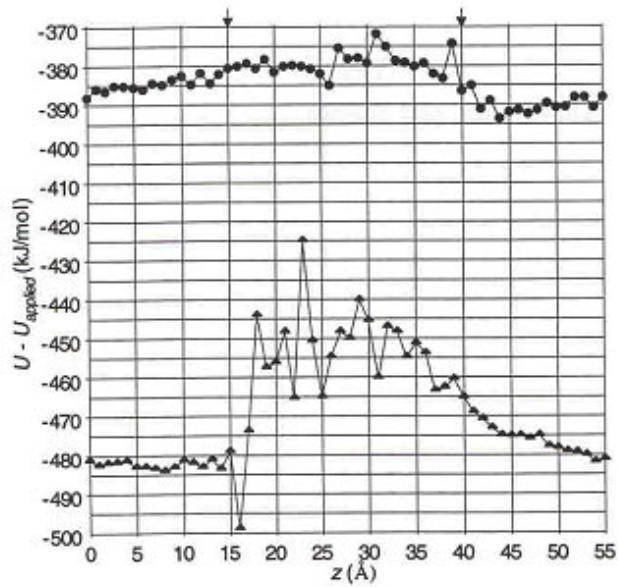

**Figure 7.** Individual ion potential energy (due to interactions with the whole system, excluding the applied potential) as a function of z for the 2 M and 0.55 V applied potential case. This figure differs from Figures 6 in that the simulated ions themselves were used to compute the potentials (no test particle insertions were used) and LJ interactions were included. Only ions within 2 Å of a line drawn through the center of the channel were included in the averaging.

*Ion potential energy decomposition*

There is enough sampling with these long trajectories that the average ion-environment interaction energy can be used directly as a rough gauge of the free energy barriers for ion passage, without resorting to specialized sampling techniques such as umbrella sampling. In Figure 7, we present one of the major components of the total system potential energy, the ion-environment interaction energy, as a function of the ion z position for the case of 2 M, 0.55 V. Here the environment is defined as all water molecules, all other ions, and all membrane and channel wall atoms, as well as their long-range images, including both local and inverse space components. Excluded from the sum are the water-water, water-channel, and water-membrane interaction energies and the applied potential. Because we used the SPC/E model for water and there are no other bonded atoms in the system, only nonbonded interaction energies play a role in the interaction energy. The energies in bulk cannot be directly compared to free energies of hydration because they neglect changes in the water energy due to dissolved charges. For $Na^+$ there is a small interaction energy barrier of about 15 kJ/mol. The barrier peaks to the right of the channel center at ~30 Å and declines quite sharply towards the exit at 40 Å. The fact that the interaction energy remains favorable throughout the channel suggests that the polar channel walls can almost fully compensate for the second shell and more distant electrostatic coordination of the ion, and is consistent with MD simulations performed by Roux and Karplus (1994) for $Na^+$ entry into the gramicidin channel. In contrast, the average $Cl^-$ - environment interaction energy has a barrier of ~30 kJ/mol. We suspect that the channel water energy would also contribute further to this peak based on geometric considerations shown below. Again, the barrier tends to be highest towards the channel exit, which, for $Cl^-$, is on the left.

The increased energy barrier to $Cl^-$ passage can be partly ascribed to the channel water molecules and their interactions with the channel wall. Figure 8 shows the average H, O, $Na^+$, and $Cl^-$ densities as a function of distance from the center of the channel for one of the trajectories. Oxygen density is seen to peak near 3 Å from the channel center with H peaks on either side. Thus, the channel walls strongly attract the channel water molecules. Both $Na^+$ and $Cl^-$, when present, stay nearer the axis than most of the channel waters, which is more problematic for $Cl^-$ than for $Na^+$ because the proximity of the water oxygens would be repulsive for $Cl^-$ but attractive for $Na^+$. The structural consequences are illustrated in Figures 9 and 10, which display the frustrated coordination of $Cl^-$ and felicitous coordination of $Na^+$, respectively, by channel waters.

**Discussion**

We have computed the current flow through a model ion channel using applied field MD. Sufficient complete passages were obtained in 100 ns simulations to provide reasonable estimates of the current for membrane potentials as low as 0.5 V with permeant ion concentrations as low as 0.5 M. Although both values are above those seen in animals, such values have been attained in lipid bilayer experiments (e.g. Andersen, 1983a). This represents considerable progress toward effective modeling of ion transport through channels under realistic conditions.

To achieve closed circuit conditions and thus prevent charge buildup during and after ion passage, we utilized periodic boundaries in the z dimension. With explicit water and ions, it is conceptually reasonable to represent the effect of metalic electrodes using a constant field throughout the system. The explicit mobile water molecules and ions should respond to the applied field to produce the expected macroscopic voltage conditions, which include little resistance in the aqueous baths and a constant field across the membrane. Although non-polarizable electrodes (such as Ag/AgCl electrodes) work on a somewhat different principle, the electrostatic effects should be comparable. We have demonstrated that the expectations about the net voltage drops are reasonable on average, even for high voltages.

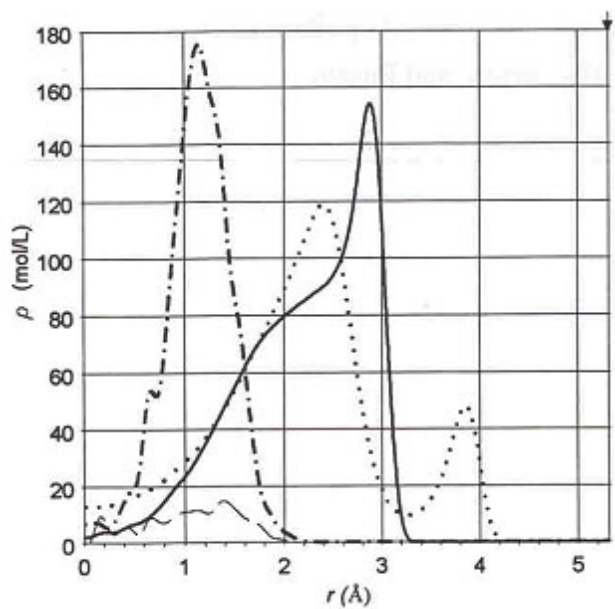

**Figure 8.** Average density of the various species inside the channel (15<z<40 Å) for the 2 M, 0.55 V case. Density is given as a function of r, the distance from the center of the channel, where r=5.3125 Å corresponds to the center of the spheres that make up the channel walls (marked by the arrow). For plotting convenience ion densities were multiplied by 100 and hydrogen densities were divided by 2. Traces for the various species are: O (solid line), H (dotted line), $Na^+$ (dash-dot line), and $Cl^-$ (dashed line).

To make simulation of multiple ion passages feasible, we minimized the system size, used a simplified channel and bilayer, used an optimized $P^3M$ algorithm, computed the trajectories for each condition piecewise in parallel, and utilized only high concentrations and voltages where the number of ions in the small baths and the driving force on the ions would be appreciable. Physiological concentrations and voltages would require much longer and larger simulations. A system with completely mobile channel and membrane atoms would obviously require much more computer processor time, especially with the addition of the nonpolar lipid atoms. However, one advantage of explicit solvation is that there is no need to employ artificial methods to represent local dielectric constants. Furthermore, our results here show that such improvements are necessary to take into account correlated motions of bath ions and water molecules that are likely to occur during ion passage. For instance, as an ion is localized to the entry of a channel, the nearby capacitative charge, channel water structure and probably the counter-ion distribution will be distorted, significantly affecting the energetics and motions of the permeating ion. In many cases, they will serve to reduce expected potential energy barriers, an expected negative feedback mechanism. These changes must be understood to correctly predict the entry rate relative to translocation and exit, which is needed to correctly assess structure-function permeation relationships, like those currently being explored with gramicidin channels (Andersen et al., 1998; Busath et al., 1998; Cotten et al., 1999; Anderson et al., in press; Thompson et al., in press; Markham et al., in press).

Here, we will first compare our apparent channel conductance and current voltage relationship shapes to those measured for biological channels (specifically the NAChR channel and gramicidin A channel). We will then compare these findings to those of recent BD simulations. Finally, we evaluate the electrical characteristics of our system and consider the statistical uncertainties and conceptual inadequacies of our approach.

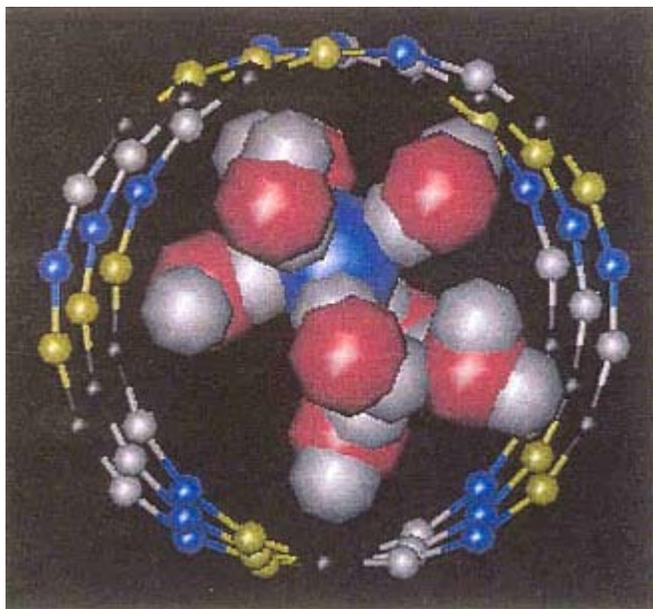

**Figure 9.** Snapshot of a chloride ion in the channel along with all water molecules and channel wall atoms that are within 3 Å (in the $z$-direction) of the chloride ion.

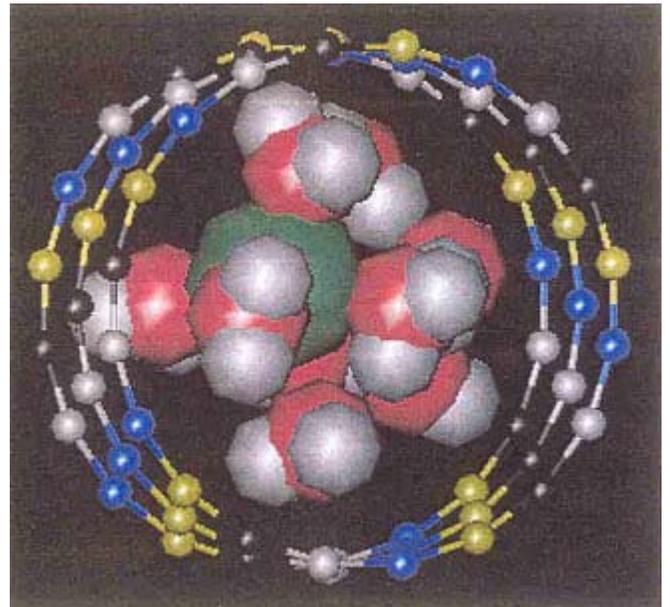

**Figure 10.** Snapshot of a sodium ion in the channel along with all water molecules and channel wall atoms that are within 3 Å (in the $z$-direction) of the sodium ion.

Our model channel is not intended to represent any specific biological protein, but has an internal diameter (8.125 Å) similar to that of the NAChR, porins, voltage-gated calcium channels, and others (cf. Hille, 1992, Koebnik et al., 2000). The wall polarity, due to the partial charges on the atoms lining the wall, was designed through the use of standard peptide unit atomic charges to be similar to that of a channel lined by the protein backbone, such as is the case in the gramicidin channel and the P region of a potassium channel, both of which are narrower in diameter and accommodate only a single file of water molecules and ions. If the channel conductance is computed in the usual fashion as the current divided by the total applied potential, a value of 32.9 pS is obtained for the 2M NaCl at 0.55 V, slightly higher than the conductance of a gramicidin A channel in 2 M NaCl, 30.4 pS at 0.2 V (Busath, et al., 1998), and lower than that of a NAChR channel, 44 and 49 pS in chromaffin cells and myotubes, respectively, under physiological conditions (as cited in Hille, 1992, pg. 332). Because the model channel used here is not structurally identical to either type of biological channel, and furthermore has rigid walls, this agreement to within an order of magnitude of experimental values is to be considered good. The model channel has somewhat lower conductance than expected for its diameter. This is probably due to the restriction of the $Na^+$ ions to the neighborhood of the channel axis by the channel waters, as well as the immobile channel walls.

The current voltage relations of gramicidin channels in phosphocholine bilayers shift from sublinear in 0.1 M NaCl to slightly superlinear in 1 M NaCl (Becker et al., 1992; Busath et al., 1998). This is consistent with the superlinearity in the model channel I-Vs for the moderate and high concentrations. The superlinearity is interesting, because at such high voltages one might expect the currents to saturate. Perhaps saturation would become evident at lower concentrations, and is beginning to appear in our 0.5 M NaCl results (although in a distorted form due to statistical fluctuations). For instance, in gramicidin channels, saturation is clearly evident when the concentration is reduced to 0.1 M (Andersen, 1983a).

The selectivity for cations observed here may be a general property of channels of the diameter used here. In our channel, water molecules were generally adsorbed on the wall surface and rotated with many of the protons pointing toward the wall and oxygen atoms near the center (Figure 8), a favorable orientation for coordination of cations but not for anions. This is reminiscent of the water structure observed for smooth-walled nonpolar 3-Å radius cylinders, suggesting that the channel water structure (and channel selectivity) may be yet a general property of the channel wall curvature (Lynden-Bell and Rasaiah, 1996).

The applied-field MD simulations can be compared to recent BD simulations. Using explicit ions and smooth-walled channels with reflective boundaries surrounding the system volume, an applied field, and induced charge at the dielectric interface boundaries, Chung and colleagues performed simulations of model NAChR (Chung et al., 1998), potassium (Chung et al., 1999), and calcium (Corry et al., 2000a; Corry et al., 2001) channels. Their simulations were typically 0.5-1.0 µs long for each state point, and they covered many more state points than we considered. It appears that their production speed was roughly 50 times faster than ours due to omission of explicit water molecules. They used a long time step of 100 fs (with a short one of 2 fs for steep regions on the energy surface) but don't include water molecules; we used a uniform time step of 2.5 fs and included explicit water molecules. Interestingly, Chung and colleagues found that the lag between the time the entry site is disoccupied and the time another ion enters (i.e. the inverse of the entry rate constant) was strikingly voltage-dependent. This is contrary to expectations from gramicidin channel experiments (Andersen, 1983a), where sublinearity of I-V relations is obtained at near-physiological cation concentrations and interpreted to imply little or no voltage-dependence of the entry step. Here, we too found a strong voltage dependence to the inverse of the mean empty time. Although the field in the conductive bath just outside the channel should be small and should not significantly affect approach of an ion to the channel, it is possible that voltage does affect the distribution of ions around the recently vacated channel or the correlated motions of ions and water molecules near the mouth of the channel.

Finally, it is interesting to relate the electrical parameters of our system to those of a macroscopic system. The effects of periodic boundaries and a small unit cell can be dramatic and need to be carefully considered to determine the legitimacy of comparisons with experimental results. In the central cell, the bilayer has an area of 625 $Å^2$, a predicted capacitance, based on membrane area, thickness and the permittivity of free space, of 0.30 $µF/cm^2$, and an apparent capacitance, based on the observed excess charge in the double layer, of 2.74 $µF/cm^2$. This derives from a net capacitive charge of $1.46 \times 10^{-19}$ C in the central cell at the 1.1 V applied potential, or 0.59 e. The excess observed membrane capacitance may be due to the channel water, which raises the effective dielectric constant of the membrane. Our observations imply that local ion concentration near a channel may be increased above bulk not only by interfacial polarization due to the average membrane capacitance (Andersen, 1983b; Becker et al., 1992), but also by a factor of up to 9 more due to the polarity of the channel and its contents. It is possible that the periodic boundary conditions result in some interactions between the central box and images that affect the ion passage. This will be explored in future calculations. Clearly, our simplified model also lacks other electrostatic features considered germane to modulation of ion channel conductance. Atomic polarizability of alkane chain atoms (which yields a typical dielectric constant of ~2) would somewhat reduce the electrostatic barrier to translocation (Jordan, 1984). The interfacial dipole layer, known to affect gramicidin channel conductance (Busath et al., 1998) is not properly represented by the water-membrane interface in the model. However, judging from the Na$^+$-environment interaction energy (Figure 7), the so-called "image barrier" is quite low in our system (~15 kJ/mol), probably due to the moderately large diameter of the channel and continued solvation of the ion in the channel. The interfacial dipole potential in our system is ~0.3 V (Figure 6A), similar to that measured for phospholipid bilayers (Pickar and Benz, 1978).

Electrophysiologists might be concerned about a possible series resistance artifact in our system. The specific conductivity of 1M NaCl is 12.9 Ω-cm (measured in our lab). For an accessible volume of 25Å length (total) and 625 Å$^2$, the resistance of the bath regions in our central cell should be 51.6 MΩ. Compared to the channel resistance, which, using the conductance for 1 M NaCl at 1.1 V in Table 2, is 44.5 GΩ, this is a negligible series resistance. We note, in this context, that the observed conductance of the channel is lower than expected for a volume conductor with bulk conductivity and the dimensions of the channel, which, from the channel's ion accessible radius (~2 Å), the channel length, and the bulk solution resistivity, would be ~390 pS in 1 M NaCl. This can be attributed to immobilization of the channel water molecules by the polar, rigid channel walls reducing ion mobility in the pore and to the potential energy barrier for ion crossing, as well as other atomistic factors.

In summary, multiple ion passages can be simulated in a model channel if sufficient concentration and voltage are utilized. The use of periodic boundary conditions prevents artificial charge buildup and yields closed circuit conditions. A uniform applied field results in appropriate compensations by SPC/E water molecules and explicit Na$^+$ and Cl$^-$ ions. Although applied-field MD simulations are currently too time-consuming to use for modeling extensive physiological data sets, it is nevertheless evident that extensive sampling of trajectories and direct computations of channel currents, albeit under extreme conditions, are feasible with this method. The results demonstrate the importance of including explicit solvent in the simulation and suggest that future analysis of ion transport trajectories can be carried out without presuppositions about the solvent response or the reaction coordinate.

**Acknowledgements**

We thank Nathan Holladay and John Harb for valuable comments and discussion on the results. This project was supported in part by the National Science Foundation (Grant No. CHE98-13729) and by the National Institutes of Health (Grant No. AI 23007).